\documentstyle[twoside]{article}

\catcode`\@=11
\long\def\@makefntext#1{
\protect\noindent \hbox to 3.2pt {\hskip-.9pt  
$^{{\eightrm\@thefnmark}}$\hfil}#1\hfill}		%CAN BE USED 

\def\@makefnmark{\hbox to 0pt{$^{\@thefnmark}$\hss}}	%ORIGINAL 
	
\def\ps@myheadings{\let\@mkboth\@gobbletwo
\def\@oddhead{\hbox{}
\rightmark\hfil\eightrm\thepage}   
\def\@oddfoot{}\def\@evenhead{\eightrm\thepage\hfil
\leftmark\hbox{}}\def\@evenfoot{}
\def\sectionmark##1{}\def\subsectionmark##1{}}

\oddsidemargin=\evensidemargin
\newcounter{sectionc}\newcounter{subsectionc}\newcounter{subsubsectionc}
\renewcommand{\section}[1] {\vspace{12pt}\addtocounter{sectionc}{1} 
\setcounter{subsectionc}{0}\setcounter{subsubsectionc}{0}\noindent 
	{\tenbf\thesectionc. #1}\par\vspace{5pt}}
\renewcommand{\subsection}[1] {\vspace{12pt}\addtocounter{subsectionc}{1} 
	\setcounter{subsubsectionc}{0}\noindent 
	{\bf\thesectionc.\thesubsectionc. {\kern1pt \bfit #1}}\par\vspace{5pt}}
\renewcommand{\subsubsection}[1] {\vspace{12pt}\addtocounter{subsubsectionc}{1}
	\noindent{\tenrm\thesectionc.\thesubsectionc.\thesubsubsectionc.
	{\kern1pt \tenit #1}}\par\vspace{5pt}}
\newcommand{\nonumsection}[1] {\vspace{12pt}\noindent{\tenbf #1}
	\par\vspace{5pt}}

\newcommand{\textlineskip}{\baselineskip=13pt}
\newcommand{\smalllineskip}{\baselineskip=10pt}

\def\eightcirc{
\begin{picture}(0,0)
\put(4.4,1.8){\circle{6.5}}
\end{picture}}
\def\eightcopyright{\eightcirc\kern2.7pt\hbox{\eightrm c}} 

\newcommand{\publisher}[2]{{\begin{center}\footnotesize\smalllineskip 
	Received #1\\
	Revised #2
	\end{center}
	}}

\def\abstracts#1#2#3{{
	\centering{\begin{minipage}{4.5in}\baselineskip=10pt\footnotesize
	\parindent=0pt #1\par 
	\parindent=15pt #2\par
	\parindent=15pt #3
	\end{minipage}}\par}} 

\renewenvironment{thebibliography}[1]
	{\frenchspacing
	 \ninerm\baselineskip=11pt
	 \begin{list}{\arabic{enumi}.}
        {\usecounter{enumi}\setlength{\parsep}{0pt}     
	 \setlength{\leftmargin 12.7pt}{\rightmargin 0pt} %FOR 1--9 ITEMS
         \setlength{\itemsep}{0pt} \settowidth
	{\labelwidth}{#1.}\sloppy}}{\end{list}}

\newcounter{itemlistc}
\newcounter{romanlistc}
\newcounter{alphlistc}
\newcounter{arabiclistc}

\def\@citex[#1]#2{\if@filesw\immediate\write\@auxout
	{\string\citation{#2}}\fi
\def\@citea{}\@cite{\@for\@citeb:=#2\do
	{\@citea\def\@citea{,}\@ifundefined
	{b@\@citeb}{{\bf ?}\@warning
	{Citation `\@citeb' on page \thepage \space undefined}}
	{\csname b@\@citeb\endcsname}}}{#1}}

\newif\if@cghi
\def\cite{\@cghitrue\@ifnextchar [{\@tempswatrue
	\@citex}{\@tempswafalse\@citex[]}}
\def\citelow{\@cghifalse\@ifnextchar [{\@tempswatrue
	\@citex}{\@tempswafalse\@citex[]}}
\def\@cite#1#2{{$\null^{#1}$\if@tempswa\typeout
	{IJCGA warning: optional citation argument 
	ignored: `#2'} \fi}}

\def\@refcitex[#1]#2{\if@filesw\immediate\write\@auxout
	{\string\citation{#2}}\fi
\def\@citea{}\@refcite{\@for\@citeb:=#2\do
	{\@citea\def\@citea{, }\@ifundefined
	{b@\@citeb}{{\bf ?}\@warning
	{Citation `\@citeb' on page \thepage \space undefined}}
	\hbox{\csname b@\@citeb\endcsname}}}{#1}}

\def\@refcite#1#2{{#1\if@tempswa\typeout
        {IJCGA warning: optional citation argument
	ignored: `#2'} \fi}}

\def\refcite{\@ifnextchar[{\@tempswatrue
	\@refcitex}{\@tempswafalse\@refcitex[]}}

\def\pmb#1{\setbox0=\hbox{#1}
	\kern-.025em\copy0\kern-\wd0
	\kern.05em\copy0\kern-\wd0
	\kern-.025em\raise.0433em\box0}

\def\fnt#1#2{\footnotetext{\kern-.3em
	{$^{\mbox{\scriptsize #1}}$}{#2}}}

\def\runninghead#1#2{\pagestyle{myheadings}
\markboth{{\protect\footnotesize\it{\quad #1}}\hfill}
{\hfill{\protect\footnotesize\it{#2\quad}}}}
\font\tenrm=cmr10
\font\tenit=cmti10 
\font\tenbf=cmbx10
\font\bfit=cmbxti10 at 10pt
\font\ninerm=cmr9

\font\eightrm=cmr8

\def\qed{\hbox{${\vcenter{\vbox{			%HOLLOW SQUARE
   \hrule height 0.4pt\hbox{\vrule width 0.4pt height 6pt
   \kern5pt\vrule width 0.4pt}\hrule height 0.4pt}}}$}}

\begin{document}

\newpage

\runninghead{L.M. Nieto, H.C. Rosu, M. Santander} 
{Hydrogen atom}

\normalsize\textlineskip
\thispagestyle{empty}
\setcounter{page}{1}

%\copyrightheading{}                     %{Vol. 0, No.0 (1992) 000--000}

\vspace*{0.88truein}

%\fpage{1} %%%%%%%%%%%%%%%%%%%%%%%%%%%%%%%%%%%%%%%%%%%%%%%%%%%%%%%%%%%
\centerline{Mod. Phys. Lett. A 14 (35) (1999) 2463-2469 
[quant-ph/9911010 v3]}
\centerline{\footnotesize\it To understand hydrogen is to understand all of 
physics}
\centerline{\footnotesize in ``The Yin and Yang of Hydrogen", D. Kleppner,
Phys. Today, April 1999, pp. 11-12}
\bigskip
\bigskip
\centerline{\bf HYDROGEN ATOM AS AN EIGENVALUE PROBLEM IN 3D}
\centerline{\bf SPACES OF CONSTANT CURVATURE AND MINIMAL LENGTH}
\vspace*{0.035truein}
%\centerline{\bf MANUSCRIPTS USING COMPUTER SOFTWARE\footnote{For
%the title, try not to use more than 3 lines. Typeset the title
%in 10 pt Times Roman, uppercase and boldface.}}
\vspace*{0.37truein}
%\centerline{\footnotesize NAME}
%\footnote{Typeset names in
%10 pt Times Roman, uppercase. Use the footnote to indicate the
%present or permanent address of the author.}}
%\vspace*{0.015truein}
%\centerline{}
%\baselineskip=10pt
%\centerline{\footnotesize\it City, State ZIP/Zone,
%Country\footnote{State completely without abbreviations, the
%affiliation and mailing address, including country. Typeset in 8
%pt Times Italic.}}
%\vspace*{10pt}
\centerline{\footnotesize  L.M. NIETO$^{1}$, H.C. ROSU$^{2}$, 
M. SANTANDER$^{1}$}
\vspace*{0.015truein}
%\centerline{\footnotesize [Received 17 August 1999] }
%\baselineskip=10pt
%\centerline{\footnotesize \it 2. Instituto de F\'{\i}sica,
%Universidad de Guanajuato, Apdo Postal E-143, 37150 Le\'on, Gto, Mexico}
\centerline{\footnotesize\it 1. Departamento de F\'{\i}sica Te\'orica, 
Universidad de Valladolid, 47011 Valladolid, Spain}
\centerline{\footnotesize \it 2. Instituto de F\'{\i}sica,
Universidad de Guanajuato, Apdo Postal E-143, 37150 Le\'on, Gto, Mexico}
\vspace*{0.225truein}
\publisher{(August 18, 1999)}{(November 3, 1999)}

%%%%%%%%%%%%%%%%%%%%%%%%%%%
\vspace*{0.21truein}
\abstracts{An old result of A.F. Stevenson 
[{\em Phys. Rev.} {\bf 59}, 842 (1941)]
concerning the Kepler-Coulomb  quantum problem on the
three-dimensional (3D) hypersphere is considered from the perspective of 
the radial Schr\"odinger
equations on 3D spaces of any (either positive, zero or negative) constant
curvature. Further to Stevenson, we show in detail how to get the
hypergeometric wavefunction for the hydrogen atom case. Finally, we make
a comparison between the ``space curvature" effects and minimal length
effects for the hydrogen spectrum.}{}{}
%%%%%%%%%%%%%%%%%%%%%%%%%%%

%\vspace*{10pt}
%\keywords{The contents of the keywords}

\textlineskip                  %) USE THIS MEASUREMENT WHEN THERE IS
\vspace*{12pt}                 %) NO SECTION HEADING

\vspace*{1pt}\textlineskip	%) USE THIS MEASUREMENT WHEN THERE IS
%\section{General Appearance}    %) A SECTION HEADING
\vspace*{-0.5pt}
\noindent

%%%%%%%%%%%%%%%%%%%%%%%%%%%%%%%%%%%%%%%%%%%%%%
%PACS number(s):  98.80.Hw, 11.30.Pb

\noindent
%%%%%%%%%%%%%%%%%%%%%%%%%%%%%%%%%%%%%%%%%%%%%%%%%%%%%%%%%%%%%%%%%%%%%

%\newpage

%\pagebreak

%\textheight=7.8truein
%\setcounter{footnote}{0}
%\renewcommand{\thefootnote}{\alph{footnote}}

%\section{The Main Text}

{\bf 1}. Ever since Schr\"odinger first considered the hydrogen atom in the 3D
hypersphere, the 3D space of constant positive curvature,\cite{S}
quantum mechanics in curved spaces has been of strong
interest due to possible astrophysical and cosmological
applications.\cite{astro}

The aim of this work is to present a discussion of the
radial Schr\"odinger problem in 3D spaces of constant curvature (either positive,
zero or negative) including explicitly the curvature parameter in the formalism,
thus being more general from the mathematical standpoint than any
study previously done and entailing early works as particular cases. In the final part
of the work we discuss possible fundamental length effects and provide a comparison
with the constant curvature one.

{\bf 2}.
We start by recalling the famous result of Schr\"odinger,\cite{S} who showed
by his factorization method that in a 3D hypersphere (the 3D space of
constant positive curvature $\kappa=1/R^2$), the eigenspectra
of the Kepler-Coulomb potential described by the radial potential 
$V(r) = -e^2/(R \tan(r/R))$ was given by:
%%%%%%%%%%%%%%%%%%%%%%%%
\begin{equation}
E_n=B\left(- \frac{1}{n^2}+(n-1)(n+1)\frac{a_1^2}{R^2}\right)~,
\end{equation}
%%%%%%%%%%%%%%%%%%%%%%%%%%%
where $B$ is Rydberg's constant, $a_1$ is Bohr's radius, and $R$ denotes
the radius of curvature of the Universe. This is a result of amazing
simplicity, where the spectra of the H-atom in flat space and the spectra of the
hypersphere itself combine {\em additively} to give the spectrum of the H-atom 
in the hypersphere. As commented by Schr\"odinger,  this formula allows for a
smooth transition from `bound' to `free' motion for
$n\approx \sqrt{R/a_1}$ a value for which Schr\"odinger's estimate is $10^{18}$, 
the crowding of the Bohr (Rydberg) states gradually going into the crowding that
represents the continuum. After the first impetus given by Schr\"odinger,
several other authors,
among whom we mention Infeld, Stevenson, Hull, Higgs, Leemon, Barut and
Wilson,\cite{auth} contributed to further mathematical clarifications of the
problem. Recently, Debergh studied the isotropic
oscillators of various dimensionalities in curved space 
by means of Witten's supersymmetric quantum mechanics,\cite{deb} whereas
Bonnor investigated the effects of cosmic expansion on the hydrogen
atom.\cite{bonn}

In a three dimensional space of constant curvature $\kappa$
the equivalent of the time-independent Schr\"odinger equation is $(-\Delta
_{LB}^{\kappa}+ U_{\kappa}(\vec{x}))\Psi _{\kappa}(\vec{x})
=\lambda  \Psi _{\kappa}(\vec{x})$, where
$\Delta _{LB}^{\kappa}=|g_{\kappa}|^{-1/2}\frac{\partial}{\partial
x_i}(|g_{\kappa}|^{1/2}g_{\kappa}^{ij}\frac{\partial}{\partial x^j})$ is the
Laplace-Beltrami
operator for the metric $(g_\kappa)_{ij}$,  $U_{\kappa}=\frac{2m}{\hbar
^{2}}V_{\kappa}$ is the potential function, and 
$\lambda  =\frac{2m}{\hbar ^{2}}E $ is the eigenvalue
parameter; the label $\kappa$ will remind we are working on the 3D space of
nonzero constant curvature $\kappa$. For ``polar" coordinates $(r, \theta, \phi)$,
the metric tensor of a space with constant curvature $\kappa$ is
$g_{\kappa}={\rm diag} (1,S_{\kappa}^2(r), S_{\kappa}^2(r)\sin ^2\theta)$, where
the {\em ``curved" sinus} $S_{\kappa}(r)$ is defined to be 
${\textstyle\frac{1}{\sqrt\kappa}\sin}(\sqrt\kappa r)$,
$r$, or 
${\textstyle\frac{1}{\sqrt{-\kappa}}\sinh}(\sqrt{-\kappa} r)$,
 for positive ($\kappa >0$), zero ($\kappa =0$), or negative
curvature ($\kappa <0$) respectively.\cite{not}
After separating the standard spherical harmonics (there is still ordinary
$SO(3)$ rotational invariance for any value of $\kappa$)
$\Psi _{\kappa}(\vec{x})= Y_l^m(\theta, \phi) G_{\kappa}(r)$, the following radial
equation can be obtained:
\begin{equation}
\left[ -\frac{1}{S_{\kappa}
^2(r)}\frac{d}{d
r}\left(S_{\kappa}^2(r) \frac{d}{d
r}\right)+\frac{l(l+1)}{S_{\kappa} ^2
(r)}+U_{\kappa}(r)
-\lambda \right] G_{\kappa}(r)=0~.
\end{equation}
This radial equation does not reduce to the standard
1D Schr\"odinger form with an effective potential but
instead still contains a first derivative term of the type 
$-\frac{2}{T_{\kappa}(r)} G_{\kappa}^{'}(r)$, where $T_{\kappa}(r)$ is a
similar notation for the curved tangent function as that for the curved
sinus, so that 
$T_{\kappa}(r)={\textstyle\frac{1}{\sqrt\kappa}\tan}(\sqrt{\kappa} r)$ for
$\kappa>0$, 
$T_{\kappa}(r)=r$ for
$\kappa=0$, and 
$T_{\kappa}(r)={\textstyle\frac{1}{\sqrt{-\kappa}}\tanh}(\sqrt{-\kappa} r)$ for
$\kappa<0$. Elimination of this first derivative term, leading to various
so-called {\em normal} forms of  Eq~(2), can be done in many ways as exemplified
by the works on the hydrogen atom in curved space of positive
curvature. Of course, the spectrum does not depend on the method employed.

In this approach the radial coordinate $r$ has dimensions of
length. There are two length scales at hand. One  is introduced through the
coupling constant of the potential, and  in the case of the hydrogen
atom is the Bohr radius $a_1=\hbar ^2/me^2$. The other, which does not appear in
flat space, is set by the space ``curvature radius", $R=1/\sqrt{\kappa}$ or
$R=1/\sqrt{-\kappa}$ according as $\kappa>0$ or 
$\kappa<0$.

{\bf 3}.
We present now details on the (Kepler-Coulomb) hydrogen atom problem
in spaces of constant curvature. The potential is
$V(r)=-e^2/(R \tan(r/R))$ for positive and $V(r)=-e^2/(R \tanh(r/R))$ 
for negative curvature, respectively; these potentials as well as the euclidean one
can be described together as $V_\kappa(r) = -e^2/{\rm T}_{\kappa}(r)$.  Using the
change of the independent variable
$y=1/ {\rm T}_{\kappa}(r)$, the term with the first derivative disappears and the
radial equation is writen in the form
\begin{equation}
G^{''}_{\kappa}+\left[\frac{{\lambda}+\beta y}{(\kappa+y^2)^2}
-\frac{l(l+1)}{\kappa+y^2}\right] G_{\kappa}=0~,
\end{equation}
where $\beta =\frac{2me^2}{\hbar ^2}=\frac{2}{a_1}$.
For {\em positive} curvature and using ``natural" units where $\kappa =1$ and $r$ 
appear as the dimensionless angle $\chi=r/R$, measured in radians,  Eq~(3) is
precisely Stevenson's equation 2.

If $\kappa\neq 0$, Eq~(3) has three singular regular points at $y=\{\pm
\sqrt{-\kappa}, \infty\}$; when $\kappa=0$ there is an irregular singularity at
$y=0$. We are going to solve first the $\kappa\neq 0$ case and $\kappa=0$ will follow as a
limiting case. The existence of three singularities suggests us to introduce a new change of
variable in order to place the singularities at the standard positions
$\{0,1,\infty\}$; this is accomplished by means of a M\"obius transformation,
indeed $y= \sqrt{-\kappa} (z-2)/z$, leading to an equation for a function, say
$f(z)$, for which a standard decomposition of the type
$f(z)=z^{p}(1-z)^{q}g(z)$ is applied. This leads to the
equation for the function $g(z)$
\begin{equation}
z(1-z)g{''}(z)+[2+2p-2(p+q+1)z]g{'}(z)+C(z)\, g(z)=0~,
\end{equation}
where the coefficient $C(z)$ depends on $\{p,q,l,
\beta,{\kappa}, \lambda\}$ and reduces to the constant $-ab$ of  hypergeometric type
if the following two conditions are satisfied
\begin{equation}
p(p+1)=l(l+1)~,
\qquad 
q^2 -q=\frac{\lambda -\beta\sqrt{-\kappa}}{4\kappa}~.
\end{equation}
From Eqs (4,\,5), the following identification of the
hypergeometric parameters $a$, $b$ and $c$ in terms of $l$ and $p, q$ can
be obtained
\begin{equation}
a+b+1=2(p+q+1)~,
\qquad
ab= (p+1)(p+2q)- \frac{\beta\sqrt{-\kappa}}{4\kappa}~,
\qquad
c=2p+2~.
\end{equation}
The first Eq in (5) has two solutions $p=l$ and $p=-(l+1)$; our choice will be the 
first one to avoid singularities at $r=z=0$. For simplicity, let us
introduce two new parameters
\begin{equation}
\omega_+= \sqrt{\frac{\kappa+\lambda +\beta\sqrt{-\kappa}}{\kappa}}~,\qquad
\omega_-= \sqrt{\frac{\kappa+\lambda -\beta\sqrt{-\kappa}}{\kappa}}~.
\end{equation}
From the two possible solutions of the second Eq in (5) 
\begin{equation}
q_\pm=\frac{1}{2}\left(1\pm \omega_- \right)~,
\end{equation}
we choose $q_+$, and therefore the solutions of (6) are
\begin{equation}
a_+=l+1+ \frac{1}{2}\left(\omega_- -\omega_+\right)~,  \qquad
b_+=l+1+ \frac{1}{2}\left(\omega_- +\omega_+\right)~\qquad c=2l+2
\end{equation}
We have now all the ingredients to write down the solution of (4), but first remark
that in order to have a well behaved {\it physical\/} solution, the function
${}_2F_1(a_+,b_+; c;z)$ has to be a polynomial, which implies
$a_+=-m\in\{0,-1,-2,\dots\}$. Using (7) and (9), this condition leads us to the
following energy quantization
\begin{equation}
\lambda=-\frac{\beta^2}{4 n^2}+(n^2-1)\kappa\ \ {\rm or}\ \
E_n=B\left(  -\frac{1}{ n^2}+(n^2-1)\kappa a_1^2 \right), \ \
n=l+m+1,
\end{equation}
where $B=me^4/(2\hbar^2)$ is Rydberg's constant. The physical solution for $g(z)$
is  
\begin{equation}
g(z)= {}_2F_1(l+1+(\omega_- -\omega_+)/2,l+1+(\omega_- +\omega_+)/2; 2l+2;z)~,
\end{equation}
where the first argument  is a negative
integer $-m$ and the hypergeometric function reduces to a Jacobi polynomial
$P_{m}^{(2l+1,-n+(\omega_- +\omega_+)/2)}(1-2z)$. From here the solution $G_\kappa(r)$
can be found to be
\begin{eqnarray}
G _{\kappa}(r) \!&\! \propto \!&\! (S_{\kappa}( r))^{l} \ e ^{-\sqrt{-\kappa}\,
r(l+2q_+)} \times  \\
 \!&\! \!&\! \  \textstyle {}_2F_1(l+1+\frac{\omega_-
-\omega_+}2,l+1+\frac{\omega_- +\omega_+}2; 2l+2;  2\sqrt{-\kappa} \,
S_{\kappa}(r) \, e^{-\sqrt{-\kappa}\, r}) \nonumber \\
\!&\! \propto \!&\!  
(S_{\kappa}( r))^{l} \ e ^{-\sqrt{-\kappa}\, r(l+2q_+)}
P_{m}^{(2l+1,-n+(\omega_- +\omega_+)/2)}(1-4\sqrt{-\kappa} \,
S_{\kappa}(r) \, e^{-\sqrt{-\kappa}\, r}).  \nonumber 
\end{eqnarray}

{\bf 4}.  We will show  now that the flat space limit of Eq~(12) leads to the
well known result in terms of associated Laguerre polynomials. Taking
into account the following approximations for small values of $\kappa$, and the standard
limiting relation from the Jacobi to Laguerre polynomials: 
\begin{eqnarray}
 S_{\kappa}(r) \!&\! = \!&\!  r+O(\kappa)~, \\ 
\sqrt{-\kappa}\, (l+2q_+) \!&\! = \!&\!  \sqrt{-\lambda_0} r+O(\sqrt{\kappa})~, \\
\frac{\omega_- -\omega_+}2\!&\! = \!&\!  \frac{\beta}{2\sqrt{-\lambda_0}} + O(\kappa)~,  \\
 l+1+\frac{\omega_- +\omega_+}2 \!&\! = \!&\! \sqrt{\frac{-\lambda_0}{-\kappa}} +
              O(\sqrt{\kappa})~, \\
\sqrt{-\kappa} \, S_{\kappa}(r) \, e^{\sqrt{-\kappa}\, r} 
      \!&\! = \!&\!  \sqrt{-\kappa} r + O(\kappa)~, \\ [1ex]
\lim_{\nu\to 0} P_m^{(\mu, \nu)}(1-2x/\nu)  \!&\! = \!&\!  L_m^\mu (x)~,
\end{eqnarray}
where $\lambda_0=-\beta^2/(4n^2)$ is the value of the eigenvalue
parameter for $\kappa=0$. From the second form in Eq (12) we get
\begin{eqnarray}
G _{0}(r) \!&\! \propto \!&\! r^{l} e^{-\sqrt{-\lambda _{0}}r}
\lim _{\kappa\rightarrow 0} P_{m}^{(2l+1,-n+(\omega_- +\omega_+)/2)}(1-4\sqrt{-\kappa} \,
S_{\kappa}(r) \, e^{-\sqrt{-\kappa}\, r})  \nonumber\\
 \!&\! \propto \!&\!  r^{l} e^{-\sqrt{-\lambda _{0}}r}
L_m^{2l+1}(2\sqrt{-\lambda_0}r)~.
\end{eqnarray}
The same result can be of course obtained directly from the first form in Eq (12) by using
the definition of the confluent hypergeometric function ${}_{1}F_{1}(a,c;z)=\lim
_{b\rightarrow
\infty}{}_{2}F_{1}(a,b,c;\frac{z}{b})$, and the fact that when the first argument
is a negative integer $-m$ this reduces to a Laguerre polynomial, 
${}_{1}F_{1}(-m,2l+2;z)\propto L_{m}^{2l+1}(z)$. 
Therefore  we reach the
right flat space result, starting from either the 3D spherical  or 3D hyperbolic spaces.

{\bf 5}. At the request of the Editor we briefly discuss whether or not recently
proposed  ``minimal length" effects may carry significance
for the H-atom.\cite{ur1} These effects lead to
modifications of the uncertainty relations and are related to either
vogue ideas implied by string theories and non-commutative geometry,
or by any type of possible non-pointlike structures within any quantum
particle,\cite{ur2} in our case the electron.
At the formal level, a rather general procedure of
modifying the uncertainty principles is provided by
the Kempf-Mangano-Mann\cite{ur3} change of the mixed commutator in the
Heisenberg algebra
\begin{equation}
[\hat{x}_i,\hat{p}_{j}]=i\hbar(\delta _{ij}+\gamma \hat{p}_i\hat{p}_j+
\gamma \delta _{ij}\hat{p}^2)
\end{equation}
where $\sqrt{\gamma}={\cal L}/\hbar$ is a deformation parameter introducing the
new length scale ${\cal L}$ in the problem at hand. If gravitational effects
are taken into account there are minute modifications in other
commutators as well.\cite{ahluw}

In the first perturbative order in $\gamma$ the consequences
of this `minimally' modified Heisenberg algebra on the H-spectrum
have been recently studied by Brau.\cite{ur4} We write Brau's result
for the corrected energy eigenvalues in the slightly changed form
\begin{equation}
E_{n,l}=B\left(-\frac{1}{n^2}+4\left(\frac{{\cal L}}{a_1}\right)^2
\frac{4n-3(l+1/2)}{n^4(l+1/2)}\right)~.
\end{equation}
As one can see, the correction is always positive and is maximal for
the ground state, leading
to a relative decrease $Q_{{\cal L}}=\Delta E_1/E_1=-20({\cal L}/a_1)^2$ of 
the hydrogen ionization energy, which would mean a very tiny 
calibration effect in the Rydberg
constant. Within each $n$ multiplet the effect is maximal for the $l=0$ levels:
$\Delta E_{n,0}/E_{n,0}= - 4({\cal L}/a_1)^2(8n-3)/n^2$. This suggests
looking to the accuracy in the frequency data for the $1S-2S$
transition, which at the present time is 1KHz.\cite{ur5} The precision in
the energy difference
between the two levels is about $10^{-12}$ eV, implying
${\cal L}\leq 0.01$ fm. If we discard any non-pointlike structure within the
electron and claim that a fundamental length scale parameter may be only the
Planck length, ${\cal L}=L_{P}$, then the relative effect in the Rydberg
calibration constant is $Q_{P}=-20 (L_{P}/a_1)^2
\approx -2\cdot 10^{-48}$.

Putting together in the H-spectrum
the extremely small fundamental effects discussed here we get the following
formula
%%%%%%%%%%%%%%%%%%%%%%%%
\begin{equation}
E_{n,l}=B\left(- \frac{1}{n^2}+4\left(\frac{{\cal L}}{a_1}\right)^2
\frac{4n-3(l+1/2)}{n^4(l+1/2)}+ \kappa (a_1)^2 (n^2-1)\right)~.
\end{equation}
%%%%%%%%%%%%%%%%%%%%%%%%%%%
For the ground state, $n=1$,
the constant curvature (or ``Schr\"odinger") effect does not
contribute and one is left with the minimal length effect (for the Planck case
of relative order $10^{-48}$ as aforementioned).
For $n=2$, and for the estimate $R/a_1 \approx 10^{36}$, the cosmological effect
which may be either positive or negative, according to the sign of the space
curvature, is of the order
$|\Delta E_2/E_2|\approx 10^{-71}$, whereas the order of a Planckian effect
is $\Delta E_2/E_2\approx 10^{-48}$. The two effects become comparable (of relative order
$10^{-52}-10^{-53}$)
around $n=10^5$, but this is a Rydberg region as yet unavailable, not
to mention the ``Schr\"odinger" range $n=10^{18}$ where the ``cosmological" effect
is of relative order one. The present day detected Rydberg atoms are at $n\leq 10^3$.
However, all these hopeless estimates change drastically
for ${\cal L}\gg L_{P}$ and/or in strong local curvature fields.

%\newpage

\nonumsection{Acknowledgements}

\noindent
This work was supported by Projects from CONACyT (No.~458100-5-25844E), DGES
from the Spanish Ministerio de Educaci\'on (PB98-0370), and Junta de Casti\-lla y
Le\'on (CO2/97). H.C.R. wishes  to acknowledge the kind hospitality at the
Departamento de F\'{\i}sica Te\'orica, U. de Valladolid, and F. Brau for
correspondence. We acknowledge D.V. Ahluwalia for suggesting to comment on the
minimal length effect as related to the context of this work,  and for other useful
remarks.

\end{document}